\documentclass[a4paper,11pt]{article}
\pdfoutput=1 % to allow pdflatex compilation in JCAP
\usepackage{jcappub}
\usepackage{amsmath}
\usepackage{graphicx}
\usepackage{comment}
\usepackage{latexsym}
\usepackage{xspace}
\usepackage{lscape}
\usepackage{color}
\usepackage{hyperref}
\usepackage{bm}
\usepackage{relsize}
\usepackage{tabularx}
\usepackage{multirow}
\usepackage{amssymb}
\usepackage[table]{xcolor}
\newcommand{\ee}[1]{\begin{equation}#1\end{equation}}
\newcommand{\ea}[1]{\begin{align}#1\end{align}}

\providecommand{\f}[2]{\frac{{#1}}{{#2}}}
\def\baq{\begin{eqnarray}}
\def\eaq{\end{eqnarray}}

\makeatletter
\gdef\@fpheader{}
\g@addto@macro\bfseries{\boldmath}
\makeatother

\newcommand{\dd}{\mathrm{d}}
\newcommand{\efold}{$e$-fold}

\newcommand{\beq}{\begin{equation}}
\newcommand{\eeq}{\end{equation}}
\newcommand{\bea}{\begin{eqnarray}}
\newcommand{\eea}{\end{eqnarray}}

\newlength{\wsingfig}
\newlength{\wdblefig}
\newlength{\wquadfig}
\newlength{\wtriplefig}
\title{Renormalisation group improvement \\ in the stochastic formalism}

\author[a,b]{Robert J. Hardwick,}
\author[c,d]{Tommi Markkanen,}
\author[e]{Sami Nurmi\,}

\affiliation[a]{Institute of Cosmology \& Gravitation, University of Portsmouth, Dennis Sciama Building, Burnaby Road, Portsmouth, PO1 3FX, United Kingdom}
\affiliation[b]{Department of Infectious Disease Epidemiology, Faculty of Medicine, Imperial College London, South Kensington Campus, London SW7 2AZ, United Kingdom}
\affiliation[c]{Department of Physics, Imperial College London, Blackett Laboratory, London, SW7 2AZ, United Kingdom}
\affiliation[d]{Laboratory of High Energy and Computational Physics, National Institute of Chemical Physics and
Biophysics, R\"avala pst. 10, Tallinn, 10143, Estonia}
\affiliation[e]{Department of Physics, University of Jyv\"{a}skyl\"{a}, P.O. Box 35, FI-40014 University of
Jyv\"{a}skyl\"{a}, Finland}

\emailAdd{r.hardwick@imperial.ac.uk}
\emailAdd{t.markkanen@imperial.ac.uk}
\emailAdd{tommi.markkanen@kbfi.ee}
\emailAdd{sami.t.nurmi@jyu.fi}

\abstract{We investigate compatibility between the stochastic infrared (IR) resummation of light test fields on inflationary spacetimes and renormalisation group running of the ultraviolet (UV) physics. Using the Wilsonian approach, we derive improved stochastic Langevin and Fokker-Planck equations which consistently include the renormalisation group effects. With the exception of stationary solutions, these differ from the naive approach of simply replacing the classical potential in the standard stochastic equations with the renormalisation group improved potential. Using this new formalism, we exemplify the IR dynamics with the Yukawa theory during inflation, illustrating the differences between the consistent implementation of the UV running and the naive approach.

}


\begin{document}
\begin{flushleft}
	\hfill		  IMPERIAL/TP/2019/TM/02
\end{flushleft}
\sloppy

%\arxivnumber{16XX.XXXXX}

\maketitle

\section{Introduction}
\label{sec:intro}

A period of early cosmic inflation provides solutions to several cosmological conundrums~\cite{Starobinsky:1980te, Sato:1980yn, Guth:1980zm, Linde:1981mu, Albrecht:1982wi, Linde:1983gd}. It is well known that perturbatively computed correlators of light scalar fields exhibit logarithmic infrared (IR) divergences during inflation. This is due to the secular growth of long-wavelength modes~\cite{Weinberg:2005vy,Seery:2010kh}, sourced by quantum fluctuations continuously crossing the horizon. The infrared issues are closely connected to the non-existence of a physical propagator for a free, minimally coupled massless scalar in de Sitter space~\cite{Allen:1985ux,Allen:1987tz} and several non-perturbative resummation techniques for the interactive cases have been developed %{\bf [SN: Can we reformulate this? The logic is not fully clear: no free propagator  $\rightarrow$ must resum diagrams which contain propagators.]}  
 e.g., references~\cite{Serreau:2011fu,Prokopec:2011ms,Arai:2012sh,Garbrecht:2011gu,Burgess:2009bs,Herranen:2013raa}. %%{ \bf (Removed a few sentences TM)}

The stochastic formalism~\cite{Starobinsky:1982ee, Starobinsky:1986fx, Starobinsky:1994bd} is a widely used method for investigating the IR dynamics in the exponentially squeezed quantum state during inflation~\cite{PhysRevLett.59.2555,
PhysRevD.46.1440,PhysRevD.42.3413}. It is an approximative coarse-grained formalism where the impact of subhorizon fluctuations is modeled as stochastic noise.  The dynamics of the long-wavelength field $\bar{\phi}$ is governed by the Langevin equation 
\cite{Starobinsky:1982ee, Starobinsky:1986fx, Starobinsky:1994bd}
\begin{equation} \label{eq:langevin}
\frac{\partial }{\partial t}\bar{\phi} ({\bf x},t) = - \frac{1}{3H}\frac{\partial V{}}{\partial \bar{\phi}}  + f ({\bf x},t) ~.
\end{equation}
Here $f({\bf x},t)$ is the stochastic source term with correlation properties of white noise and the amplitude set by the solution of the linearised mode equation of the quantum field. The stochastic approach effectively performs a re-summation rendering the (quantum) correlators free of IR divergences. In specific examples it has been shown to correctly reproduce the leading-log order IR results derived by other means~\cite{Tokuda:2017fdh,Guilleux:2015pma,Guilleux:2016oqv,Arai:2011dd,Moreau:2018ena,Moreau:2018lmz,Prokopec:2017vxx}. Recent works addressing foundations of the stochastic approach include references~\cite{Moss:2016uix,Rigopoulos:2016oko,Tokuda:2018eqs,Cruces:2018cvq,Glavan:2017jye,Hardwick:2017fjo,Vennin:2015hra,Grain:2017dqa,Firouzjahi:2018vet}.

During inflation light energetically subdominant scalars probe field values from the vacuum parametrically up to the Hubble scale $H$. Quantum corrections may induce significant running of couplings over this window and must be accounted in studying the dynamics. This feature in conjunction with the measured central values for couplings of the Standard Model (SM) of particle physics pointing to a metastable electroweak vacuum imply that Higgs fluctuations could easily have triggered a fatal transition to the true minimum in the early Universe~\cite{Degrassi:2012ry}, see reference~\cite{Markkanen:2018pdo} for a recent review. The fact that this did not happen can be used to place a stringent constraint on the Higgs non-minimal coupling~\cite{Herranen:2014cua,Herranen:2015ima,Markkanen:2018bfx}, which is the last unknown parameter in the SM, and further can be used as a novel test to constrain SM extensions. 

In quantum field theory, the couplings of a Lagrangian depend on the renormalisation scale $\mu$. The choice of the scale is arbitrary and has no effect on physical quantities, which translates into a set of renormalisation group equations, which can be solved to obtain the renormalisation group improved theory. This renormalisation group improved theory corresponds to a resummation of infinite classes of perturbative corrections and therefore remains valid over a larger range of scales compared to the standard loop expansion at the same order in perturbation theory. 

It is not immediately obvious if the renormalisation group techniques can be applied as such in the approximative stochastic formalism, where the field is split by hand into a quantum UV part and a classical IR part, the UV part is substituted by the linearised solution of mode equations and the decaying quantum modes are dropped. Indeed, as we will discuss below, a simple replacement of the classical potential by the renormalisation group improved effective potential $V_{\rm eff}(\phi)$ in the stochastic Langevin equation~\eqref{eq:langevin} is not consistent in general. The field correlators computed in this way fail to obey the renormalisation group equations, i.e. the Callan-Symanzik (CS) equations~\cite{Symanzik:1970rt,Callan:1970yg}
\beq
\label{eq:CZforphin}
\frac{{\rm d}\langle\bar{\phi}^n\rangle}{{\rm d}{\rm ln}\mu} = - n\gamma \langle\bar{\phi}^n\rangle~,
\eeq 
except in the limit of stationary solutions (here $\mu$ is the renormalisation scale and $\gamma$ is the field anomalous dimension). A violation of the CS equations is a sign of inconsistency, signalling a $\mu$ dependence of quantities which should be measurable. However, we will show that this problem is resolved when also the renormalisation of the kinetic term is correctly accounted for in the Langevin equation. Starting from the Wilsonian approach to the renormalisation group, we provide a reformulation of the stochastic formalism which is manifestly consistent with the renormalisation group improvement. We then move on to apply the formalism in the specific example of a Yukawa theory and discuss the qualitative effects of the renormalisation group corrections in this case. For earlier works addressing the Yukawa theory stochastically or renormalization group improvement during inflation see \cite{Miao:2006pn} and \cite{Bilandzic:2007nb}, respectively.  

The paper is organised as follows. In section~\ref{sec:Lange} we briefly review the Wilsonian approach to renormalisation group and rewrite the stochastic Langevin equation into a form compatible with the renormalisation group improvement. In section~\ref{sec:FP} we investigate the corresponding Fokker-Planck equation and show that the correlation functions satisfy the correct CS equations. In section~\ref{sec:examples} we study the magnitude of RG effects in the stochastic formalism by considering the example of Yukawa theory on curved spacetime and, finally, in section~\ref{sec:conclusions} we conclude with a summary of our findings and prospects for future applications. Our sign conventions are $(-,-,-)$ according to reference~\cite{Misner:1974qy}.

\section{The Renormalisation Group and the Langevin equation}
\label{sec:Lange}
\subsection{Wilsonian Renormalization}

The standard formulation of the stochastic approach see, e.g., references~\cite{Garbrecht:2013coa,Hardwick:2017fjo,Starobinsky:1994bd}, starts from the operator equation of motion of the form $(\Box+V')\hat{\phi} = 0$. The field is split into IR and UV parts and the UV part is substituted with the linearised solution. Dropping the decaying mode of the UV part, the approximative equation of motion for the IR part takes the form of a Langevin equation. Here we repeat the same procedure accounting for quantum corrections which in general affect both the kinetic term and the potential. 

%In \cite{Starobinsky:1994bd} it was shown how a Langevin equation for scalar field fluctuations in an inflationary space is obtained starting from the tree-level equation of motion  for the field operator by splitting the field into IR and UV parts and substituting the solution of the linear mode equations for the UV part. The same idea can be applied also beyond the tree-level. 

To set up the framework, we follow the Wilsonian approach and introduce a hierarchy of scales 
\beq
\label{murange}
H < \mu < \Lambda~,
\eeq 
where $\Lambda$ is a UV cutoff and $H$ is the Hubble rate. The Wilsonian coarse-graining scale $\mu$ can be effectively thought as the renormalisation scale $\mu$ introduced when computing quantum corrections with an infinite cutoff $\Lambda\rightarrow \infty$. 

Integrating out modes in the momentum shell $\mu < k < \Lambda$ the generating functional takes the form 
\beq
Z[J] = \underset{\hspace{-15 pt}k_{\rm E} < \mu}{\int {\cal D}\phi} \underset{\hspace{-0 pt}\mu < k_{\rm E} < \Lambda}{\int {\cal D}\phi}\, e^{-\int \dd^4 x_{\rm E} \,[{\cal L}_{\rm E}(\phi)+J\phi \,]} =  \underset{\hspace{-15 pt}k_{\rm E} < \mu}{\int {\cal D}\phi} \,  e^{i\int \dd^4 x \,[{\cal L}_{\mu}(\phi) +J_{\mu}\phi \,]}\,,
\eeq 
where ${\cal L}_{\rm E}$ is the Euclidean action and ${\cal L}_{\mu}$ is the effective action of the coarse-grained theory describing modes with a cut-off in Euclidean momentum space $k_{\rm E} < \mu$. From now on we will drop the subscript ``${\rm E}$'' and it is to be understood that the cut-off is defined in Euclidean space.

Neglecting terms with more than two derivatives, the effective action is given by the standard expression~\cite{Peskin:1995ev}
\beq
\label{L(mu)}
{\cal L}_{\mu} = \frac{1}{2}\nabla^{\nu}(Z^{1/2}(\mu)\phi(\mu))\nabla_{\nu}(Z^{1/2}(\mu)\phi(\mu)) - V_{\rm eff}(\mu,\phi(\mu),\{\lambda_i(\mu)\})~.  
\eeq
The field operator $\phi(\mu)$ takes the form
\beq
\phi(\mu) = Z^{-1/2}(\mu) \phi(\Lambda)~,\label{eq:fieldr}
\eeq
where the factor $Z$ is due to the wave function renormalisation. Choosing  $Z(\Lambda) = 1$, it is given by 
\beq
\label{Z}
Z(\mu) = {\rm exp}\left[{\int_{\Lambda}^{\mu}2\gamma(\mu') \,\dd{\rm ln} \mu'}\right]~,
\eeq
where $\gamma$ is the anomalous field dimension defined by 
\beq
\label{gamma}
\frac{{\rm d} \phi(\mu) }{{\rm d}{\rm ln}\mu} = -\gamma\phi(\mu)~.
\eeq

Throughout this work, we will ignore all possible field dependent contributions in $Z(\mu)$, which is often called the local potential approximation \cite{Guilleux:2016oqv}. Since our approach only includes the very UV part in the effective action (in contrast to what is usually done) with the IR calculated stochastically this is expected to be a good approximation: the high UV physics does not suffer from secular IR effects, and can be addressed via standard perturbative quantum field in theory in curved space, where such terms have little impact, especially when close to de Sitter space where kinetic contributions are naturally suppressed. 

The effective potential $V_{\rm eff}$ in (\ref{L(mu)}) contains in principle all quantum corrections, but as mentioned at the UV limit where only the modes $\mu < k <\Lambda $ are integrated over. Since $H < \mu$, the modes are at least marginally subhorizon. Any contribution from IR sensitive terms in $V_{\rm eff}$ should be small in this limit and is neglected here. We approximate the build-up of IR effects entirely using the stochastic approach. 

The UV effective potential $V_{\rm eff}$ obeys the Callan-Symanzik equation given by 
\beq
\label{CZforV}
\frac{{\rm d} V_{\rm eff}}{{\rm d}{\rm ln}\mu} = \left(-\gamma\phi\frac{\partial}{\partial \phi} + \sum_i\beta_i\frac{\partial}{\partial \lambda_i}+ \frac{\partial}{\partial {\rm ln}\mu}\right)V_{\rm eff}= 0~, 
\eeq
where $\lambda_i$ denote couplings of the theory and the beta functions are defined as usual $\beta_i = {\rm d} \lambda_i/{\rm d}{\rm ln}\mu$. The scaling relations (\ref{Z}), (\ref{gamma}) and (\ref{CZforV}) ensure that the effective action (\ref{L(mu)}) does not depend on the choice of the renormalisation scale $\mu$. 

The quantum equation of motion for the field operator $\phi$ is obtained by varying the Lagrangian (\ref{L(mu)}), which yields 
\beq
\label{EOMchifull}
Z(\mu) { \Box} \phi(\mu) + V'_{\rm eff}(\mu,\phi(\mu),\lambda_i(\mu)) = 0~,
\eeq
where the prime denotes derivative with respect to the field $\phi(\mu)$.  This quantum corrected equation is our starting point for setting up the stochastic formalism and studying its compatibility with the renormalisation group improvement.

For reference, we list here the remaining Callan-Symanzik equations that will be needed in our analysis below. The derivatives of the effective potential scale as  
\beq
\label{CZforVn}
\frac{{\rm d} V^{(n)}_{\rm eff}}{{\rm d}{\rm ln}\mu} = n\gamma V^{(n)}_{\rm eff}~.
\eeq
From the $\mu$ dependence of $\phi(\mu)$ given in (\ref{eq:fieldr}) we can write the CS equation for an $n$-point function as
\begin{align}
\frac{\dd}{\dd \mu}\left[ \big( \sqrt{Z} \big)^{n}\langle\phi(x_1)\phi(x_2)\cdots\phi(x_n)\rangle\right]&=0\nonumber \\ \Leftrightarrow\quad\bigg( \mu \frac{\partial}{\partial \mu}+\beta_{\lambda_i} \frac{\partial}{\partial \lambda_i}+n\gamma\bigg) \langle\phi(x_1)\phi(x_2)\cdots\phi(x_n)\rangle&=0\,\label{eq:CZ0}.
\end{align} 
Here, and in what follows, we suspend explicit notation of the full set of arguments whenever this does not compromise definiteness.

\subsection{Running in the Langevin equation}

Next, we follow precisely the standard steps in setting up the stochastic approach~\cite{Garbrecht:2013coa,Hardwick:2017fjo,Starobinsky:1994bd} but use the full quantum corrected equation of motion (\ref{EOMchifull}) with an arbitrary renormalisation scale $\mu$.  We split the 
field operator into IR and UV parts with a sharp cutoff defined by a coarse graining parameter $\sigma\lesssim 1$
\beq
\label{chisplit}
\phi = \int \frac{{\rm d} {\bf k}}{(2\pi)^3}\theta(\sigma a H-\vert {\bf k}\vert )\phi_{\bf k}e^{i {\bf k}\cdot{\bf x}}+\int \frac{{\rm d} {\bf k}}{(2\pi)^3}\theta(\vert {\bf k}\vert -\sigma a H) \phi_{\bf k}e^{i {\bf k}\cdot{\bf x}}  \equiv \bar{\phi} +\varphi\,.
\eeq 
It should be noted that the coarse graining scale $k/a = \sigma H$ is always below the renormalisation scale $\mu$ which throughout the work is chosen to lie in the UV window (\ref{murange}). 

Substituting (\ref{chisplit}) into the equation of motion (\ref{EOMchifull}) and expanding in the UV field $\varphi$ yields  
\beq
\label{0}
Z{ \Box} \bar{\phi} + V'_{\rm eff}(\bar{\phi}) + \left[Z { \Box}+ V''_{\rm eff}(\bar{\phi}) \right]\varphi + {\cal O}(\varphi^2)= 0~. 
\eeq
Following reference~\cite{Starobinsky:1994bd}, we substitute the UV part $\varphi$ with  
\beq 
\phi_{\bf k} = a_{\bf k} u_{\bf k} + a^{\dag}_{-\bf k} u^{*}_{-\bf k}\,,
\eeq
where the annihilation and creation operators satisfy the usual commutation relations and the mode functions $u_{\bf k}(t,\mu)$ are determined by the linearised mode equation 
\beq 
\ddot{u}_{\bf k}+ 3H \dot{u}_{\bf k}-\frac{\vert {\bf k}\vert^2}{a^2}u_{\bf k} +  V''_{\rm eff} Z^{-1}u_{\bf k}  = 0~.
\eeq
Note that the effective mass term $V''_{\rm eff} (\mu,\bar{\phi}(\mu))/ Z(\mu)$ does not depend on the RG scale since the $\mu$ depencies of $V''_{\rm eff}$ and $Z(\mu)$ precisely cancel.  Concentrating on the limit of light fields $V''_{\rm eff} \ll H^2$ and choosing the Bunch-Davies vacuum, the mode functions are just the usual Hankel functions\footnote{The origin of the $Z^{-1/2}$ factor here follows directly from the scaling of the 2-point function in \eqref{eq:CZ0}.}
\beq 
\label{chiuvHankel}
u_{\bf k}(\tau,\mu) = Z^{-1/2}(\mu) (-\tau)^{3/2}H(\tau)\frac{\sqrt{\pi}}{2}(1-\epsilon)H_{\nu}^{(1)}(-\vert {\bf k}\vert \tau)~.
\eeq
Here ${\rm d}\tau = {\rm d} t/a$ defines the conformal time, $H_{\nu}^{(1)}$ is the Hankel functon of the first kind, and the index $\nu$ is determined by 
\beq
\nu = \sqrt{\frac{9}{4}+3(\epsilon -\eta Z^{-1})}~,\qquad \epsilon = -\frac{\dot{H}}{H^2}~,\qquad \eta= \frac{V''_{\rm eff}}{3H^2}~.
\eeq 

Next one substitutes the linear solution (\ref{chiuvHankel}) for $\varphi$ back to the full equation (\ref{EOMchifull}) and drops terms  $ {\cal O}(\varphi^2)$~\cite{Starobinsky:1994bd}.  Dropping also the IR term $ \ddot{\bar{\phi}}$ which is subdominant on the slow roll attractor, one obtains  
\beq
\label{QMlangevin}
\dot{\bar{\phi}} +  \frac{V_{\rm eff}'}{3 H Z}  = \sigma aH^2 (1-\epsilon)\int \frac{{\rm d} {\bf k}}{(2\pi)^3}\delta(\vert {\bf k}\vert -\sigma a H)\phi_{\bf k}\, e^{i {\bf k}\cdot {\bf x}}  + {\cal O} (\eta \dot{\varphi})~.
\eeq
This is still an operator equation where $\bar{\phi}$ and $\varphi$ are quantum fields. However, the growing mode of $\dot{\varphi}$ commutes with the field $\varphi$ on superhorizon scales and the quantum state becomes exponentially squeezed~\cite{Albrecht:1992kf,Grishchuk:1990bj}.   
Therefore, choosing $\sigma \lesssim 1$ and neglecting the decaying mode, one can replace~\eqref{QMlangevin} with a classical, stochastic Langevin equation 
\beq
\label{langevin}
\dot{\bar{\phi}} +  \frac{V_{\rm eff}'}{3 H Z} = G^{1/2}\xi~.
\eeq
As usual, the IR field $\bar{\phi}$ is a classical stochastic variable, and the UV source behaves as white noise, hence $\langle\xi(t) \rangle = 0\,,$ $\langle\xi(t)\xi(t')\rangle = \delta(t-t')$ and the amplitude is  
\baq
\label{noise}
\ G(t, \bar{\phi}(\mu),\mu) &=& H(1-\epsilon)\frac{|u_{\bf k}|^2\vert {\bf k}\vert^3}{2\pi^2} \left. \rule{0pt}{3ex}\right |_{\vert {\bf k}\vert = \sigma aH} \nonumber \\
&\simeq& Z^{-1}(\mu)\frac{H^3}{4\pi^2}\left[ 1+2\left(\frac{\eta(\bar{\phi}(\mu),\{\lambda_i(\mu)\},\mu)}{Z(\mu)}-\epsilon\right){\rm ln}\sigma\right]~.
\eaq

The equation (\ref{langevin}) is the quantum corrected version of the standard Langevin equation in the stochastic formalism. It holds for any choice of the renormalisation scale $\mu$ which can be verified by a direct computation. Indeed, from~\eqref{gamma} and the definition of the IR field in~\eqref{chisplit}, it follows that 
\beq
\label{dphidmu}
\frac{{\rm d} \bar{\phi}(\mu) }{{\rm d}{\rm ln}\mu} = -\gamma\bar{\phi}(\mu)~.
\eeq
Using this together with~\eqref{Z} and~\eqref{CZforVn}, and differentiating~\eqref{langevin} with respect to $\mu$, we get 
\beq
\label{dlangevindmu}
\frac{{\rm d} }{{\rm d}{\rm ln}\mu}\left(\dot{\bar{\phi}} +  \frac{V_{\rm eff}'}{3 H Z} - G^{1/2}\xi\right) 
= -\gamma G^{1/2}\xi -\frac{{\rm d} G^{1/2}} {{\rm d}{\rm ln}\mu} \xi = 0~.
\eeq
In the last step we used the scaling relation 
\beq
\label{dGdmu}
\frac{{\rm d}{\rm ln} G} {{\rm d}{\rm ln}\mu} =-2\gamma~,
\eeq 
which follows from~\eqref{noise}. The RG scale independence (\ref{dlangevindmu}) derives directly from the quantum equation of motion (\ref{EOMchifull}) which is our starting point and by construction independent of the RG scale.

It is now obvious that the renormalisation group improvement can be implemented in the Langevin equation (\ref{langevin}) using the standard methods as we will show in section~\ref{sec:examples}. In the next section, we write down the corresponding Fokker-Planck equation for the probability distribution and discuss its properties.

\section{Renormalisation scale invariance of the Fokker-Planck equation}
\label{sec:FP}

The IR field $\bar{\phi}(\mu)$, obeying the Langevin equation (\ref{langevin}), is a classical stochastic quantity whose correlation functions are determined by its probability distribution $P(\bar{\phi})$ through  
\beq
\label{phimucorrelator}
\langle\bar{\phi}(\mu)^n\rangle = \frac{\int {\rm d}\bar{\phi}(\mu)\; \bar{\phi}(\mu)^n P(\bar{\phi}(\mu),\{\lambda_i(\mu)\},\mu,t)}{\int {\rm d}\bar{\phi}(\mu)\; P(\bar{\phi}(\mu),\{\lambda_i(\mu)\},\mu,t)}~.
\eeq  
From~\eqref{dphidmu} and~\eqref{Z}, changing $\mu \rightarrow \tilde{\mu}$ scales the IR field as $\bar{\phi}(\tilde{\mu}) = [Z(\tilde{\mu})/Z(\mu)]^{-1/2}\bar{\phi}(\mu)$ so that 
\beq
\label{phimutilde}
\langle\bar{\phi}(\tilde{\mu})^n\rangle = \left[ \frac{Z(\tilde{\mu})}{Z(\mu)}\right]^{-n/2}\langle \bar{\phi}(\mu)^n\rangle~.
\eeq
Differentiating this with respect to the renormalisation scale $\tilde{\mu}$, one directly obtains the usual Callan-Symanzik equations for the field correlator (\ref{eq:CZ0}). 

On the other hand, we can equally write 
\baq
\label{phimutildecorrelator}
\langle\bar{\phi}(\tilde{\mu})^n\rangle &=& \frac{\left[\frac{Z(\tilde{\mu})}{Z(\mu)}\right]^{-(n+1)/2} \int {\rm d}\bar{\phi}(\mu)\; \bar{\phi}(\mu)^n P(\bar{\phi}(\tilde{\mu}),\{\lambda_i(\tilde{\mu})\},\tilde{\mu},t)}{\left[\frac{Z(\tilde{\mu})}{Z(\mu)}\right]^{-1/2} \int  {\rm d}\bar{\phi}(\mu)\; P(\bar{\phi}(\tilde{\mu}),\{\lambda_i(\tilde{\mu})\},\tilde{\mu},t)}\\\nonumber
&=&\left[\frac{Z(\tilde{\mu})}{Z(\mu)}\right]^{-n/2} \frac{ \int {\rm d}\bar{\phi}(\mu)\; \bar{\phi}(\mu)^n P(\bar{\phi}(\tilde{\mu}),\{\lambda_i(\tilde{\mu})\},\tilde{\mu},t)}{\int  {\rm d}\bar{\phi}(\mu)\; P(\bar{\phi}(\tilde{\mu}),\{\lambda_i(\tilde{\mu})\},\tilde{\mu},t)}~.
\eaq 
Comparing this to~\eqref{phimutilde} and~\eqref{phimucorrelator} we find  that they agree only if
\beq
\label{dPdmu}
P(\bar{\phi}(\mu),\{\lambda_i(\mu)\},\mu,t) = P(\bar{\phi}(\tilde{\mu}),\{\lambda_i(\tilde{\mu})\},\tilde{\mu},t)\;\; \Leftrightarrow \;\;
\frac{{\rm d} P} {{\rm d}{\rm ln}\mu} = 0~.
\eeq
This is similar to the RG scale invariance of the full QFT generating functional. It is merely a consequence of the requirement that we are free to renormalise the theory at any chosen scale and the choice does not affect the physical solutions. 

The probability distribution that corresponds to the Langevin process~\eqref{langevin} obeys a Fokker-Planck equation which in the It\^o interpretation\footnote{One can confirm that the corresponding Stratonovich process will follow the same argumentation.} reads  
\beq
\label{FPIto}
\frac{\partial P}{\partial t} = \frac{1}{3H}\frac{\partial}{\partial \bar{\phi}}\left(\frac{V_{\rm eff}'}{Z} P\right)+ \frac{1}{2}\frac{\partial^2 }{\partial \bar{\phi}^2}\left(GP\right)~.
\eeq
In order to be consistent with the condition (\ref{dPdmu}), the $\mu$ dependent entries in the Fokker-Planck equation must precisely cancel such that (\ref{FPIto}) holds for any choice of the RG scale. This is similar to the RG scale independence of the Langevin equation (\ref{dlangevindmu}) and again directly follows from our starting point  (\ref{EOMchifull}). However, to be fully explicit let us check this by direct computation.  

To this end, we differentiate separately each term of the Fokker-Planck equation (\ref{FPIto}) with respect to $\mu$. Because $\phi$ is a test field, it does not affect the time evolution of the spacetime and consequently $t$ and $H$ have no dependence of the RG scale $\mu$.  The time derivative therefore commutes with the $\mu$ derivative, and using~\eqref{dPdmu} we get 
\beq
\frac{{\rm d}}{{\rm d}{\rm ln}\mu}\frac{\partial P}{\partial t}  = \frac{\partial}{\partial t}\frac{{\rm d} P}{{\rm d}{\rm ln}\mu}  = 0~. \eeq 
To compute the $\mu$ derivatives of the other two terms, we will repetitiously apply the relation 
\beq
\label{chainrule}
\frac{\dd}{\dd{\rm ln}\mu}\frac{\partial}{\partial\bar{\phi}} = \left(-\gamma\bar{\phi}\frac{\partial}{\partial \bar{\phi}} + \sum_i\beta_i\frac{\partial}{\partial \lambda_i}+ \frac{\partial}{\partial {\rm ln}\mu}\right)\frac{\partial}{\partial\bar{\phi}} = \frac{\partial}{\partial\bar{\phi}}\frac{\dd}{\dd{\rm ln}\mu}+\gamma \frac{\partial}{\partial\bar{\phi}}~. 
\eeq
Using~\eqref{Z},~\eqref{CZforV} and~\eqref{dPdmu}, we then readily find 
\baq
\frac{{\rm d}}{{\rm d}{\rm ln}\mu}\frac{\partial}{\partial \bar{\phi}}\left(\frac{V_{\rm eff}'}{Z} P\right)&=& \frac{\partial}{\partial \bar{\phi}} \frac{{\rm d}}{{\rm d}{\rm ln}\mu}\left(\frac{V_{\rm eff}'}{Z} P\right)+\gamma\frac{\partial}{\partial \bar{\phi}}\left(\frac{V_{\rm eff}'}{Z} P\right)\\\nonumber 
&=&\frac{\partial}{\partial \bar{\phi}}\left(\gamma \frac{V_{\rm eff}'}{Z} P- 2\gamma \frac{V_{\rm eff}'}{Z} P\right)+\gamma\frac{\partial}{\partial \bar{\phi}}\left(\frac{V_{\rm eff}'}{Z} P\right) = 0\,.
\eaq
In the same way, using also~\eqref{dGdmu}, we arrive at
\baq
\frac{{\rm d}}{{\rm d}{\rm ln}\mu}\frac{\partial^2}{\partial \bar{\phi}^2}\left(GP\right)&=& \frac{\partial^2}{\partial \bar{\phi}^2} \frac{{\rm d}}{{\rm d}{\rm ln}\mu}\left(GP\right)+2\gamma\frac{\partial^2}{\partial \bar{\phi}^2}\left(GP\right)\\\nonumber 
&=&\frac{\partial^2}{\partial \bar{\phi}^2}\left(-2\gamma GP \right)+2\gamma\frac{\partial^2}{\partial \bar{\phi}^2}\left(GP\right) = 0\,.
\eaq
Combining the results, we find that 
\beq
\frac{{\rm d}}{{\rm d}{\rm ln}\mu}\left[ \frac{\partial P}{\partial t} - \frac{1}{3H}\frac{\partial}{\partial \bar{\phi}}\left(\frac{V_{\rm eff}'}{Z} P\right)- \frac{1}{2}\frac{\partial^2 }{\partial \bar{\phi}^2}\left(GP\right)\right]= 0\,,
\eeq
verifying that the Fokker-Planck equation holds true for any choice of the RG scale and complies with the condition (\ref{dPdmu}).  This is the main result of our work. 

We reiterate that the renormalisation scale $\mu$ denotes a UV scale in the window (\ref{murange}) which is always above the coarse graining scale of the stochastic approach. We have approximated all IR physics by the stochastic approach formulated with the UV limit of the effective action and the corresponding quantum equation of motion (\ref{EOMchifull}) as the starting point. This should be distinguished from e.g. \cite{Guilleux:2015pma,Guilleux:2016oqv,Arai:2011dd,Moreau:2018ena,Moreau:2018lmz,Prokopec:2017vxx} where functional renormalisation group techniques are used to gradually integrate over IR modes and scaling relations with respect to the IR cutoff are investigated. Our approach does not offer any information about this IR scaling but the RG scaling in this work refers to deep ultraviolet scaling only. 

Note that to maintain the correct RG scaling it was necessary to include the wave-function renormalisation $Z(\mu)$ (where the flow is normalized to start at $Z(\Lambda)=1$) in the drift terms of~\eqref{langevin} and~\eqref{FPIto}. Neglecting this and retaining only the running effective potential $V_{\rm eff}(\mu)$ would not be consistent with the RG scaling and correlators computed this way would fail to satisfy the Callan-Symanzik equations (\ref{eq:CZ0}). However,  
in the special case of stationary limit in strict de Sitter space, $\partial P/\partial t = 0$ and $H=const.$, the solution of the Fokker-Planck equation (\ref{FPIto}) takes the form 
\beq
\label{stationarysol}
P_{\rm stat.}(\bar{\phi}) = C {\rm exp}\left(-\frac{2}{3H}\int {\rm d}\bar{\phi}\frac{V_{\rm eff}'}{Z G}\right)~,
\eeq
where the normalisation constant $C$ does not depend on the RG scale $\mu$. Since $G(\mu)\propto Z(\mu)^{-1}$ according to equation (\ref{noise}), the wavefunction renormalisation $Z$ drops out from the equilibrium result. Hence, if one is interested only in the stationary limit, the correct RG scaling is obtained by dropping the explicit $Z$ dependence from the Fokker-Planck equation (\ref{FPIto}) and choosing $G=H^3/(4\pi^2)$, as done for example in references~\cite{Espinosa:2015qea,Herranen:2014cua,Espinosa:2007qp}. Away from the stationary limit however, the solution will depend on $Z$. This is especially important when not in strict de Sitter space, as then the stationary solution (\ref{stationarysol}) does not necessarily coincide with the late time limit \cite{Prokopec:2015owa,Cho:2015pwa}. In the next section we investigate Yukawa theory as a specific example and discuss quantitatively the effect of the $1/Z(\mu )$ for the two-point function. 

\section{Example of Yukawa theory}\label{sec:examples}

In this section we demonstrate the application of the renormalisation group improved stochastic formalism. As a specific example, we investigate a Yukawa theory with an energetically subdominant real scalar $\phi$ which can we treat as a test field in a classical background spacetime. We numerically solve for the two-point function of $\phi$ during slow roll inflation with a quadratic inflaton potential. %{\rm SN: CHECK}. Correct!

\subsection{Yukawa theory in curved spacetime}

In section~\ref{sec:Lange} we defined the effective potential $V_{\rm eff}$ in the Langevin equation (\ref{langevin}) as the Wilsonian integral over modes $H < k < \Lambda$, where $\Lambda$ is the UV cutoff. The included modes are at least marginally subhorizon and contributions from IR sensitive terms in $V_{\rm eff}$ should therefore be small. Here we neglect all the IR terms altogether and include only the deep UV part of $V_{\rm eff}$. Indeed, this is the very idea of the approximative stochastic approach.

We follow reference~\cite{Markkanen:2018pdo} and compute the UV part of the curved space effective potential using the resummed Heat Kernel approach~\cite{Parker:1984dj,Jack:1985mw} which is essentially an expansion around the local limit in configuration space. The UV expansion captures the curved space contributions in the renormalisation group running of the effective potential. { This is not a Wilsonian approach but rather: in this expansion, no IR effects are included, allowing one to extend the momentum integration down to $k=0$ and make use of dimensional regularization when calculating $V_{\rm eff}$.} The method can be straightforwardly applied  for the slow roll solution with a quadratic inflaton potential. However, the difference compared to de Sitter solution is quantitatively irrelevant within the precision of one-loop investigation and withing the limited {\efold} range we will concentrate on. For simplicity, therefore, we compute the UV effective potential of the test field $\phi$ in a de Sitter space neglecting the time variation of the Hubble rate.

The matter part of the action for our Yukawa example is given by 
\ee{S =\int \dd^4x\sqrt{|g|}\,\bigg[\f{1}{2}\nabla_\mu\phi\nabla^\mu\phi-\f{1}{2}m^2\phi^2-\f{\xi}{2}R\phi^2-\f{\lambda}{4}\phi^4+i\bar \psi\nabla\!\!\!\!/\psi-g\phi\bar\psi\psi\bigg]\,.\label{eq:yu2}}
Here $\phi$ is a real scalar field and the Dirac spinor $\psi$ contains $N_{\rm f}$ internal degrees of freedom --- { essentially leading to $N_{\rm f}$ copies of the same theory} --- and $R =12H^2$ is the Ricci curvature scalar. The non-minimal curvature coupling $\xi \phi^2 R$ has a non-trivial renormalisation group running and therefore must be included in the action. 

Quantising the action (\ref{eq:yu2}) in a classical de Sitter background unavoidably generates the gravitational operators~\cite{ParkerToms,Birrell:1982ix}
\ea{S_{g}&=-\int \dd^4x\,\sqrt{|g|}\bigg[V_{\Lambda}-\kappa R+\alpha_{1} R^2+\alpha_{2} R_{\mu\nu}R^{\mu\nu}+\alpha_{3} R_{\mu\nu\delta\eta}R^{\mu\nu\delta\eta}\bigg]\nonumber \\ &\overset{\rm dS}{\equiv}-\int \dd^4x\,\sqrt{|g|}\bigg[V_{\Lambda}-\kappa R+\alpha H^4\bigg]\label{eq:treecurve}\,,}
where $V_{\Lambda}$ is a constant and $\kappa$ and $\alpha_i$ are dimensionless couplings. In the second step we used that for the de Sitter solution the Ricci and Riemann tensors are given by 
\ee{R^2=144H^4\,\qquad R_{\mu\nu}R_{\mu\nu}=36H^4\,,\qquad R_{\mu\nu\delta\eta}R^{\mu\nu\delta\eta}=24H^4\,,}
and we have defined the common coupling $\alpha$ of the $\mathcal{O}(H^4)$ operators as 
\ee{\alpha\equiv144\alpha_1+36\alpha_2+24\alpha_3\,.\label{eq:alp}} The gravitational backreaction of (\ref{eq:treecurve}) is small and we will neglect it. However, this part will contribute to the effective potential through the $\phi$ dependent loop logarithms. 
 
The UV limit of the de Sitter effective potential for the theory  (\ref{eq:yu2}) at one loop level takes the form\footnote{The leading quantum correction to the kinetic term is subdominant as one may see e.g. from Eq. (3.2) of \cite{Kirsten:1993jn}, in accordance with our discussion in section \ref{sec:Lange}.}~\cite{Markkanen:2018pdo}
\ee{
\label{Yukawaeffpot}
V_{\rm eff}(\phi)=\f{m^2}{2}\phi^2 +6{\xi}H^2\phi^2 + \f{\lambda}{4}\phi^4+V_\Lambda-12\kappa H^2+\alpha H^4+V^{(1)}_\phi(\phi)+N_{\rm f}V^{(1)}_\psi(\phi)\,,}
where 
\ee{V^{(1)}_\phi(\phi)=%\f{1}{2}m^2\phi^2+\f{\xi}{2}R\phi^2+\f{\lambda}{4}\phi^4+\Lambda+\kappa R+\alpha_1 R^2+\alpha_2 R_{\mu\nu}R^{\mu\nu}+\alpha_3 R_{\mu\nu\delta\eta}R^{\mu\nu\delta\eta}\label{eq:treecurve} \\ &+
\f{\mathcal{M}_\phi^4}{64\pi^2}\bigg[\ln \bigg(\f{|\mathcal{M}_\phi^2|}{\mu^2}\bigg)-\f{3}{2}\bigg]-\f{\f{1}{15}H^4}{64\pi^2}\ln\bigg(\f{|\mathcal{M}_\phi^2|}{\mu^2}\bigg)%\equiv V^{(0)}(\chi)+V^{(1)}(\chi)
\label{eq:curve3}\,,}
and
\ea{V^{(1)}_\psi(\phi)=%\f{1}{2}m^2\phi^2+\f{\xi}{2}R\phi^2+\f{\lambda}{4}\phi^4+\Lambda+\kappa R+\alpha_1 R^2+\alpha_2 R_{\mu\nu}R^{\mu\nu}+\alpha_3 R_{\mu\nu\delta\eta}R^{\mu\nu\delta\eta}\label{eq:treecurve} \\ &+
-\f{4\mathcal{M}_\psi^4}{64\pi^2}\bigg[\ln\bigg(\f{|\mathcal{M}_\psi^2|}{\mu^2}\bigg)-\f{3}{2}\bigg]+\f{\f{38}{15}H^4}{64\pi^2}\ln\bigg(\f{|\mathcal{M}_\psi^2|}{\mu^2}\bigg)%\equiv V^{(0)}(\chi)+V^{(1)}(\chi)
\label{eq:curve4}\,.}
The effective masses, $\mathcal{M}^2_\phi$ and $\mathcal{M}^2_\psi$, are given by 
\ee{\mathcal{M}^2_\phi\equiv m^2+3\lambda\phi^2+\left(\xi- \f{1}{6}\right)12H^2\,,\qquad \mathcal{M}^2_\psi\equiv g^2\phi^2+H^2\,.\label{eq:effm2}} The couplings $m$, $\lambda$, $g$, $\xi$, $V_\Lambda$, $\kappa$ and $\alpha$ denote renormalised quantities with the renormalisation conditions set at the scale $\mu$. 

The effective potential does not depend on the renormalisation scale $\mu$ and obeys the Callan-Symanzik (\ref{CZforV}) equation 
\begin{align}
\frac{{\rm d} V_{\rm eff}}{{\rm d}{\rm ln}\mu}=\bigg(\mu \frac{\partial}{\partial \mu}+\beta_{m^2}\frac{\partial}{\partial m^2}+\beta_\lambda \frac{\partial}{\partial \lambda}+\beta_\xi\frac{\partial}{\partial \xi}+\beta_{V_{\Lambda}} \frac{\partial}{\partial {V_{\Lambda}}}+\beta_{\kappa}\frac{\partial}{\partial \kappa}+\beta_{\alpha}\frac{\partial}{\partial \alpha}-\gamma\phi \frac{\partial}{\partial\phi}\bigg)V_{\rm eff}=0\,.\label{eq:CZV2}
\end{align}
The anomalous dimension $\gamma$ and the $\beta$ functions are given by   
\begin{align}
16\pi^2 \gamma &=2N_{\rm f}g^2 \label{Yukawagamma}\\
16\pi^2 \beta_{m^2} &= m^2\left(6\lambda + 4N_{\rm f}g^2\right)\label{eq:la0}  \\
16\pi^2 \beta_{\lambda} &= 18 \lambda ^2+8N_{\rm f} g^2 \lambda-8N_{\rm f} g^4\label{eq:la} \\
\ 16\pi^2 \beta_{g} &=5g^3\,\label{eq:g}\\
16\pi^2 \beta_{\xi} &= \left(\xi - \f{1}{6}
\right)\left(6\lambda + 4N_{\rm f}g^2\right)\,,\label{eq:2}
\end{align}
and 
\ea{{16\pi^2}\beta_{V_\Lambda}&=\f{m^4}{2}\,\label{eq:betaA}\\ 
{16\pi^2}\beta_\kappa&=-m^2\left( \xi-\f{1}{6}\right)\, \label{eq:betaB}\\ 
{16\pi^2}\beta_{\alpha}&= 72\left( \xi-\f{1}{6}\right)^2-\f{11N_{\rm f} +1}{15}\,.%\,\\ {16\pi^2}\beta_{\alpha_2}&=\f{4N_{\rm f}-1}{180}\,\\{16\pi^2}\beta_{\alpha_3}&=\f{7N_{\rm f}+2}{40}\,, 
\label{eq:betaC}}
For a different parametrization see, e.g., references~\cite{Bando:1992wy,Elizalde:1993qh,Geyer:1996kg,Geyer:1996wg}. From the above one may see that non-zero $\xi$ and $\lambda$ in~\eqref{eq:la} and~\eqref{eq:2} are unavoidable for a non-zero $g$: they are generated by running even if at some scale we set $\xi=0=\lambda$. 

In the following we will for simplicity set $m=0$.  This allows us to neglect all dimensionful couplings of the theory since, as seen in equations~(\ref{eq:la0}), (\ref{eq:betaA}) and (\ref{eq:betaB}), $m=0, V_{\Lambda} =0, \kappa = 0$ is a fixed point of the renormalisation group flow.

In equation (\ref{Yukawaeffpot}) the couplings are evaluated at a fixed renormalisation scale $\mu$; their implicit $\mu$-dependence the $\mu$ dependence in the logarithms to satisfy (\ref{CZforV}).  The renormalisation group improved effective potential is obtained by solving for the running couplings and field $\phi(\mu)$ from equations~(\ref{Yukawagamma}) -  (\ref{eq:betaC}) and substituting them back into the effective potential ~\eqref{eq:CZV2}. This corresponds to resumming infinite sets of diagrams computed with a fixed $\mu$ and leads to improved convergence of the result~\cite{Coleman:1973jx,Kastening:1991gv}. 

We will make use of renormalisation group improvement in the sense of reference~\cite{Ford:1992mv} by including the further step where the renormalisation scale $\mu$ is chosen to minimise the loop terms (\ref{eq:curve3}) and (\ref{eq:curve4}) over a range of values $\phi$ and $R$ so that the convergence of the one loop result is optimised, see reference~\cite{Markkanen:2018bfx} for further discussion of the optimal scale choice in curved spacetime.  Denoting the optimal choice by $\mu_*$, the renormalisation group improved potential (for $m=0,V_{\Lambda} =0, \kappa = 0$) is given by 
%This feature may be used to ones advantage by choosing an optimized $\mu_*$ leading to better control of the quantum correction, see \Ref{Markkanen:2018pdo} for more details. The potential after introducing $\mu$-optimization, the renormalization group improved potential, is then
\ee{V_{\rm RGI}(\phi)={\xi(\mu_*)}6H^2\phi(\mu_*)^2 + \f{\lambda(\mu_*)}{4}\phi(\mu_*)^4+\alpha(\mu_*) H^4+ V^{(1)}_\phi(\phi(\mu_*),\mu_*)+ N_{\rm f}V^{(1)}_\psi(\phi(\mu_*),\mu_*) \,.\label{eq:effiSM}}
The running field $\phi(\mu_*)$ can be written as   
\beq
\phi(\mu_{*}) = \frac{Z(\mu_0)^{1/2}}{Z(\mu_*)^{1/2}} \phi(\mu_0) = e^{-\int_{\mu_0}^{\mu_*} {\rm d} \mu \gamma(\mu)/{\mu}}\phi(\mu_0)~,
\eeq
where $\mu_0$ is a fixed reference scale at which we define the input values of the renormalised quantities. The wave function renormalisation factor $Z(\mu)$ is defined according to equation~(\ref{Z}) and $Z(\Lambda) = 1$. 

{In the following as the input for our analysis we will choose quantities renormalised at some reference scale $\mu_0$ in a frame with a canonical kinetic term. For this it is convenient to re-scale the field as $Z^{1/2}(\mu_0)\phi(\mu)=\tilde{\phi}(\mu)$. At the scale $\mu_0$ this will lead to a canonical kinetic term $\f12(\partial\tilde{\phi})^2$ %in the action with the effective potential
%\ee{V_{\rm eff}(\tilde{\phi})={\tilde{\xi}(\mu)}6H^2\tilde{\phi}(\mu)^2 + \f{\tilde{\lambda}(\mu)}{4}\tilde{\phi}(\mu)^4+\tilde{\alpha}(\mu) H^4+ \tilde{V}^{(1)}_\phi(\tilde{\phi}(\mu),\mu)+ N_{\rm f}\tilde{V}^{(1)}_\psi(\tilde{\phi}(\mu),\mu) \,,\label{eq:effiSM0}}
and an effective potential that is a function of $\tilde{\phi}(\mu)$ and the scaled couplings
\beq
\label{inputvalues0}
\tilde{\alpha}(\mu) \equiv \alpha(\mu)~,\qquad \tilde{\lambda}(\mu) \equiv \f{\lambda(\mu)}{Z^2(\mu_0)} ~,\qquad \tilde{\xi}(\mu)\equiv\f{\xi(\mu)}{Z(\mu_0)}~,\qquad \tilde{g}(\mu)\equiv \f{g(\mu)}{Z^{1/2}(\mu_0)}~,
\eeq
with the input values for the couplings set as
\beq
\label{inputvalues}
{\alpha}(\mu_0) \equiv \alpha_0~,\qquad\f{\lambda(\mu_0)}{Z^2(\mu_0)} \equiv \lambda_0 ~,\qquad \f{\xi(\mu_0)}{Z(\mu_0)}\equiv\xi_0~,\qquad \f{g(\mu_0)}{Z^{1/2}(\mu_0)}\equiv g_0~.
\eeq
With the initial conditions as above the value for $Z(\mu_0)$ drops out, or rather, is absorbed in the initial condition in the frame with the re-scaled field $\tilde{\phi}$. Effectively, one may then perform the entire analysis with the unscaled field and simply set $Z(\mu_0)=1$ everywhere.} Explicit solutions for the running couplings are given in the appendix \ref{sec:analytic-runnings}. 

We define the running scale $\mu_*(\phi, H)$ of the renormalisation group improved potential (\ref{eq:effiSM}) by imposing the condition 
\begin{align}
&\ln \mu_*=\nonumber \\ & \quad \frac{45\left(4N_{\rm f}{\cal M}_\psi^4-{\cal M}_\phi^4\right)+(30{\cal M}_\phi^4-4H^4)\ln {\cal M}_\phi^2+4N_{\rm f}(19H^4-30{\cal M}_\psi^4)\ln {\cal M}_\psi^2}{(19N_{\rm f}-1)8H^4+60({\cal M}_\phi^4-4N_{\rm f}{\cal M}_\psi^4)} \left. \rule{0pt}{5ex}\right |_{\mu_0} \,,  \label{eq:opt-scale}
\end{align}
where all the couplings on the right hand side are evaluated at the scale $\mu_0$. This serves to approximately minimise the one-loop logarithms in (\ref{eq:effiSM}).\footnote{Evaluating the RHS of (\ref{eq:opt-scale}) at the scale $\mu_0$ amounts to neglecting the running from the loop correction, which is a next-to-leading effect in the loop expansion.} If the right hand side is evaluated at $\mu_*$, equation~(\ref{eq:opt-scale}) is the condition for the loop logarithms in (\ref{eq:effiSM}) to vanish exactly. This choice of the optimal scale was used in reference~\cite{Markkanen:2018pdo}. Here we use equation~(\ref{eq:opt-scale}) instead because it is simpler to implement numerically.

\subsection{Numerical solution for the two-point function}

We now proceed to numerically solving for the two-point function of the test field $\phi$ using the renormalisation group improved stochastic formalism. The UV effective potential $V_{\rm eff}$ and the wavefunction renormalisation $Z(\mu)$ are given by equations~(\ref{eq:effiSM}), (\ref{Z}) and (\ref{Yukawagamma}) respectively. As the background spacetime, we choose the slow roll solution corresponding to quadratic inflation, such that
\begin{equation}
\frac{H^2}{H_{\rm end}^2} = 1 + 2(N_{\rm end}-N) \,.
\end{equation}
where $H_{\rm end} = 10^{13}{\rm GeV}$. We start computation at $N = 0$ and set $N_{\rm end } = 500$. We choose to define the input parameter values at the scale $\mu_0 = 3.2\times 10^{14}{\rm GeV}$. We have explicitly checked that the one loop terms $V^{(1)}_\phi$ and  $V^{(1)}_\psi$ in equation (\ref{eq:effiSM}) remain small throughout the computation. 

\begin{figure}[h!]
\begin{center}
\includegraphics[width=0.48\textwidth]{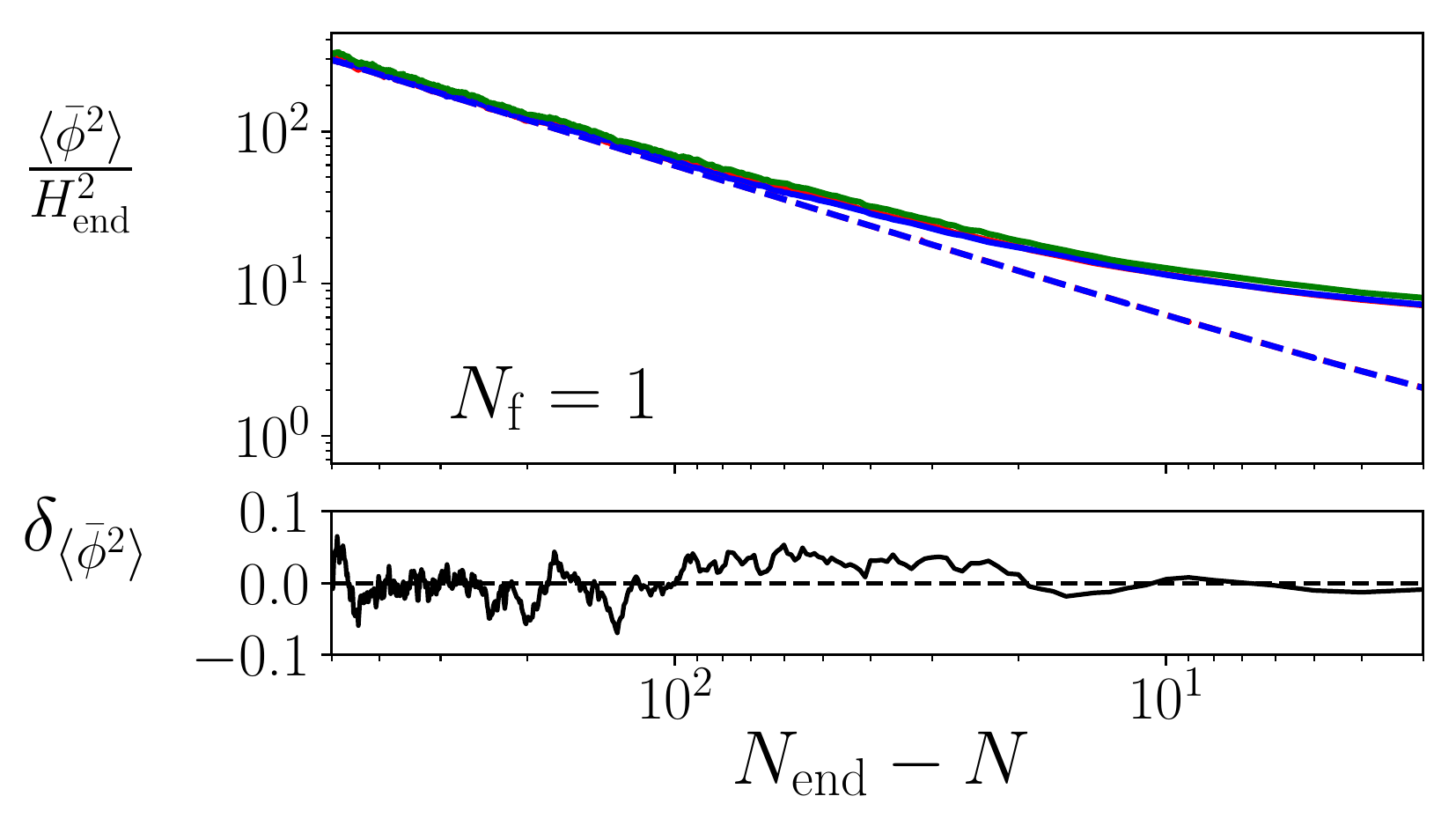}
\includegraphics[width=0.48\textwidth]{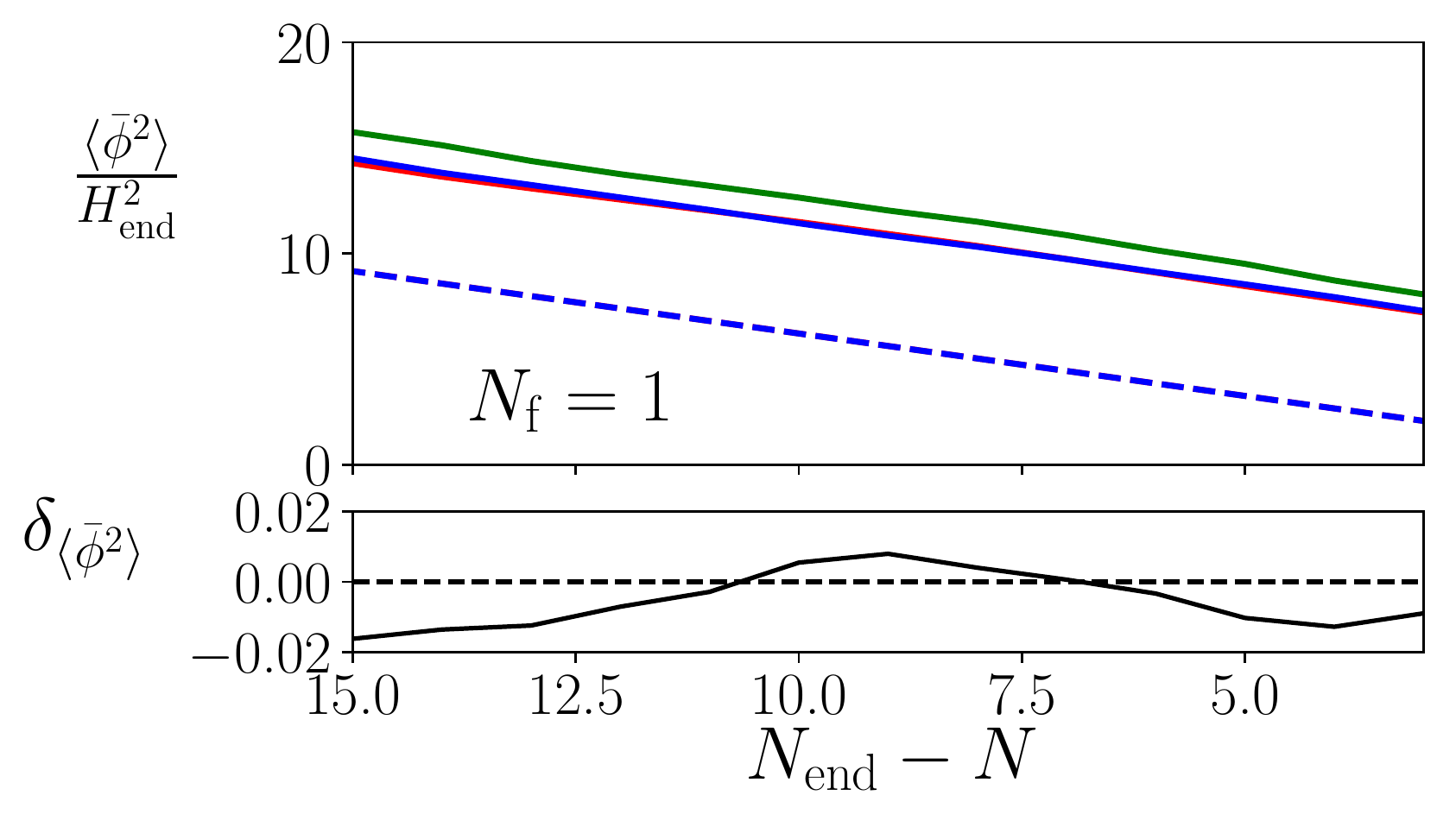} \\
\includegraphics[width=0.48\textwidth]{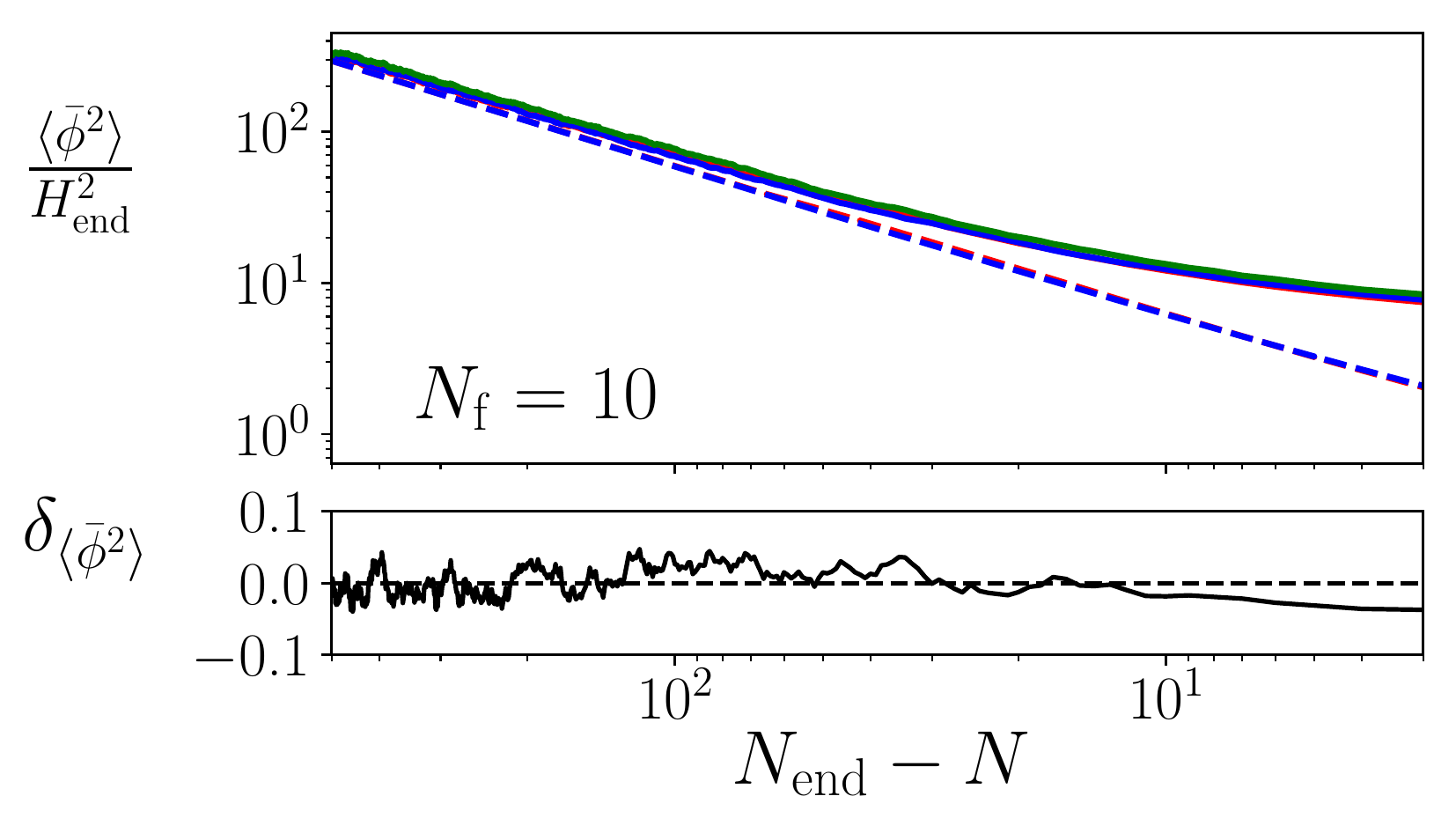}
\includegraphics[width=0.48\textwidth]{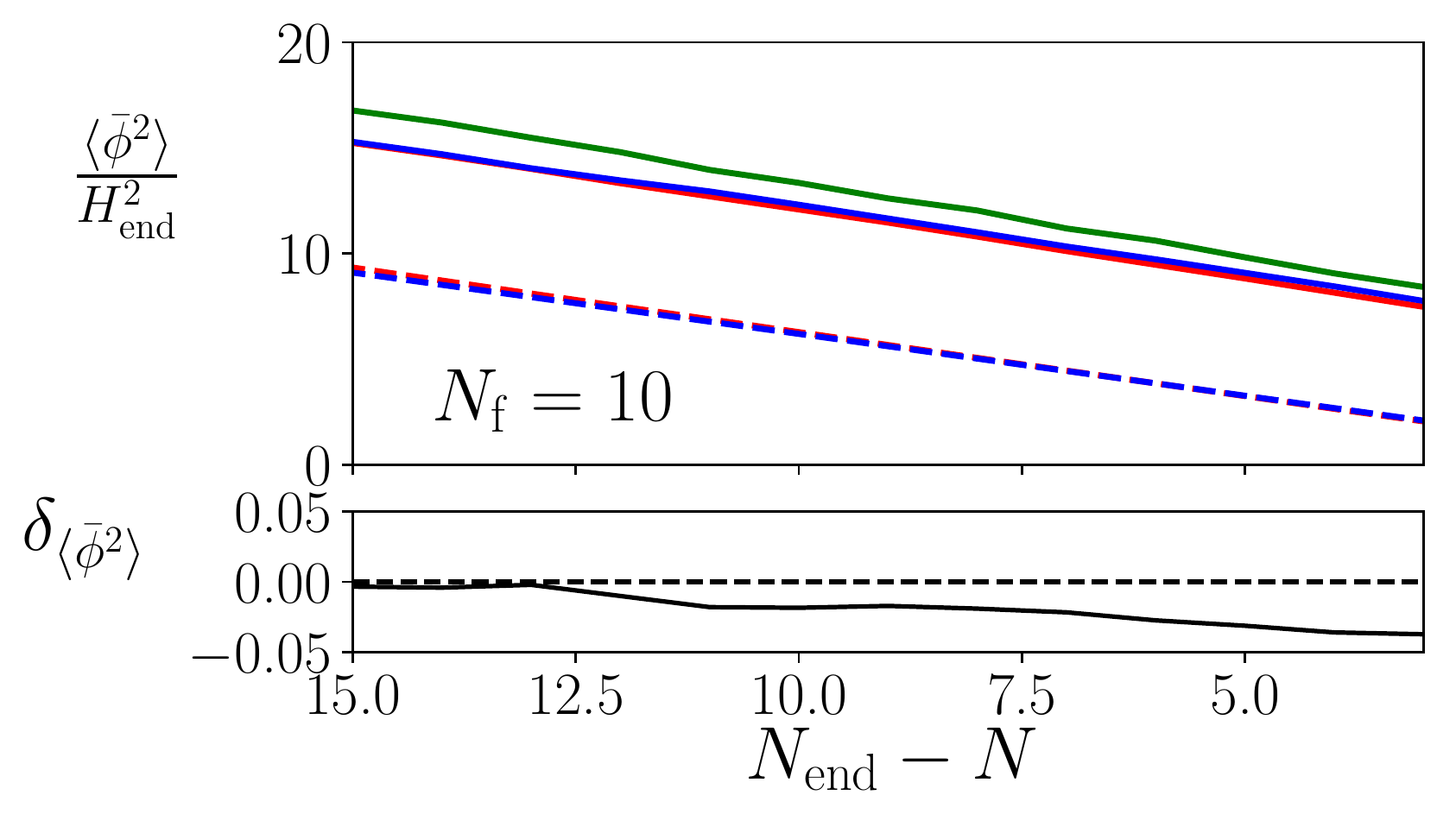}
\end{center} 
\caption{\label{fig:Yukawa-plot} The amplitude of the two-point function in Yukawa theory in a quadratic inflation background, with the following parameters used in all panels: $\xi_0=10^{-2}$; $\lambda_0=10^{-2}$; $g_0=10^{-1}$ and $m_0=0$. Solid lines denote the full numerical solution and dashed lines represent the corresponding stationary process. Each process has been numerically generated by a Langevin solver with $10^4$ realisations. Colours of the lines correspond to: the full renormalised solution for the optimal scale $\langle \bar{\phi}^2(\mu_*)\rangle$ in red, and for a fixed scale $\langle \bar{\phi}^2(\mu_0)\rangle$ in green. The naive solution with $Z$ neglected throughout is shown in blue. The fractional difference between the full and naive solutions defined in equation~(\ref{eq:delta-diff-2p}) is plotted in black below each figure. The top row plots unzoomed (left) and zoomed (right) axes setting $N_{\rm f}=1$, while the bottom row plots the equivalent with $N_{\rm f}=10$. }
\end{figure}

We determine the two-point function from a large number of numerically generated realisations of the Langevin process (\ref{langevin}). As the Langevin solver we use an adapted version of the \href{https://sites.google.com/view/nfield-py}{\texttt{nfield}} python code, publicly available at:  \href{https://github.com/umbralcalc/nfield}{https://github.com/umbralcalc/nfield}. The results are shown in figure~\ref{fig:Yukawa-plot} which depicts the time evolution of the two-point function for two parameter sets.

For comparison we have also plotted the solution of the Langevin equation (\ref{langevin}) with the wave function renormalisation factor $Z$ omitted. As discussed above, this leads to incorrect result for the two-point function and fails to satisfy the Callan-Symanzik equation (\ref{eq:CZ0}) except in the case of stationary solutions for which the probability distribution is given by equation (\ref{stationarysol}). The equilibrium solution for light spectator fields will differ from the stationary limit, depicted by the dashed line in figure~\ref{fig:Yukawa-plot}, if its relaxation time is longer than the time scale associated to the evolution of the slow roll background~\cite{Hardwick:2017fjo}.  In the figure we have also shown the difference between the full solution and the solution without the $Z$ factor defined as 
\begin{equation}
\delta_{\langle \bar{\phi}^2\rangle}\equiv \frac{\langle \bar{\phi}^2\rangle - \langle \bar{\phi}^2\rangle_{Z=1}}{\langle \bar{\phi}^2 \rangle }\,. \label{eq:delta-diff-2p} 
\end{equation}
It can be seen that the two point function $\langle \bar{\phi}^2\rangle_{Z=1}$  computed without the  $Z$ factor starts to differ from the full result $\langle \bar{\phi}^2\rangle$ as soon as the Langevin process departs from the stationary limit. The difference grows as number of fermionic fields $N_{\rm f}$ is increased, since the anomalous dimension scales proportional to $N_{\rm f}$ according to equation (\ref{Yukawagamma}). The deviation from $\langle \bar{\phi}^2\rangle$ is a signal of $\langle \bar{\phi}^2\rangle_{Z=1}$ failing to obey the RG scaling relation (\ref{eq:CZ0}).

\section{Conclusions}
\label{sec:conclusions}

In this work we have investigated how the renormalisation group running can be incorporated in the stochastic approach to inflationary infrared dynamics of light test scalars \cite{Starobinsky:1994bd}.  
%A prime example is the fate of the  Standard Model electroweak vacuum during inflation which is a delicate interplay between the renormalisation group running and the accumulation of infrared Higgs fluctuations, commonly studied using the stochastic formalism \cite{...}.  
By making use of the Wilsonian picture of renormalisation and starting from the quantum action, we have reformulated the stochastic approach in terms renormalised ultraviolet quantities with a running renormalisation scale. Our main results are equations (\ref{langevin}) and (\ref{FPIto}) which define the renormalisation group improved versions of the Langevin and Fokker-Planck equations. They differ from the corresponding standard equations in two aspects:  the classical potential is replaced with the renormalisation group improved effective potential $V_{\rm eff}$ and the field renormalisation $Z$ enters in the drift term and in the stochastic noise term. 

We have explicitly demonstrated that the renormalisation group improved stochastic approach results $n$-point functions which obey the correct Callan-Symanzik equations. The field renormalisation $Z$ plays a key role here; a simple replacement of the classical potential with $V_{\rm eff}$ in the standard stochastic equations, as was done for example in references~\cite{Espinosa:2015qea,Herranen:2014cua,Espinosa:2007qp}, is compatible with the renormalisation group scaling only in the limit of stationary solutions where $Z$ cancels out. Beyond this limit it is necessary to use (\ref{langevin}) or (\ref{FPIto}) to maintain the correct RG scaling. As shown in reference \cite{Hardwick:2017fjo}, the true equilibrium may significantly deviate from the stationary solution whenever the spacetime is not exactly de Sitter. In situations where quantum corrections and running couplings are significant, it is therefore important to use the full equations (\ref{langevin}) and (\ref{FPIto}) with the correct behaviour under the renormalisation group instead of the naive replacement. As a specific example, we have investigated the Yukawa theory in an inflationary  slow-roll background. We have numerically computed the two point function of the test scalar using the renormalisation group improved stochastic formalism and illustrated the difference compared to the naive approach with the field renormalisation term neglected. 

Two obvious situations come to mind where the renormalisation group compatible approach to the stochastic infrared dynamics is required for reliable results. The first is when investigating the Standard Model model vacuum metastability in the early universe which is fundamentally a quantum effect and sensitively depends on the RG running~\cite{Markkanen:2018pdo}. The second arises when dealing with models with a large number of fields that couple to a stochastic spectator, which can exhibit significant running and/or quantum corrections. In general, any situation involving the stochastic approach to inflation but where results beyond the simple tree-level ones are needed requires one to incorporate renormalisation group effects into the problem  which, as shown here,  can be performed in a consistent manner. Our results apply to energetically subdominant test fields but it should be straightforward to generalise the formalism to the inflaton sector as well.

\section*{Acknowledgements}

The authors would like to thank Vincent Vennin, Kimmo Kainulainen and Arttu Rajantie for useful discussions during the development of this work. RJH was supported by UK Science and Technology Facilities
Council grant ST/N5044245 throughout the duration of this work. TM is supported by the U.K. Science and Technology Facilities Council grant ST/P000762/1 and by the Estonian Research Council via the Mobilitas Plus grant MOBJD323. For efficiency, some numerical computations were performed on the Sciama High Performance
Compute (HPC) cluster which is supported by the ICG, SEPNet and the University of
Portsmouth.

\appendix

\newpage

\section{Running for the Yukawa theory in curved spacetime }
\label{sec:analytic-runnings}

It is possible to find analytic solutions for the running couplings for the theory described by (\ref{eq:yu2}) and  (\ref{eq:treecurve}). For completeness we will give the results in a general background, but for the case $m=0$. 

The one loop quantum correction on a general background can be written as~\cite{Markkanen:2018pdo}
\ee{V^{(1)}_\phi(\phi)=%\f{1}{2}m^2\phi^2+\f{\xi}{2}R\phi^2+\f{\lambda}{4}\phi^4+\Lambda+\kappa R+\alpha_1 R^2+\alpha_2 R_{\mu\nu}R^{\mu\nu}+\alpha_3 R_{\mu\nu\delta\eta}R^{\mu\nu\delta\eta}\label{eq:treecurve} \\ &+
\f{\mathcal{M}_\phi^4}{64\pi^2}\bigg[\ln \bigg(\f{|\mathcal{M}_\phi^2|}{\mu^2}\bigg)-\f{3}{2}\bigg]+\f{\f{1}{90}\left(R_{\mu\nu\delta\eta}R^{\mu\nu\delta\eta}-R_{\mu\nu}R^{\mu\nu}\right)}{64\pi^2}\ln\bigg(\f{|\mathcal{M}_\phi^2|}{\mu^2}\bigg)%\equiv V^{(0)}(\chi)+V^{(1)}(\chi)
\label{eq:curve30}\,,}
and
\ea{V^{(1)}_\psi(\phi)=%\f{1}{2}m^2\phi^2+\f{\xi}{2}R\phi^2+\f{\lambda}{4}\phi^4+\Lambda+\kappa R+\alpha_1 R^2+\alpha_2 R_{\mu\nu}R^{\mu\nu}+\alpha_3 R_{\mu\nu\delta\eta}R^{\mu\nu\delta\eta}\label{eq:treecurve} \\ &+
-\f{4\mathcal{M}_\psi^4}{64\pi^2}\bigg[\ln\bigg(\f{|\mathcal{M}_\psi^2|}{\mu^2}\bigg)-\f{3}{2}\bigg]+\f{\f{1}{90}\left(\f{7}{2}R_{\mu\nu\delta\eta}R^{\mu\nu\delta\eta}+4R_{\mu\nu}R^{\mu\nu}\right)}{64\pi^2}\ln\bigg(\f{|\mathcal{M}_\psi^2|}{\mu^2}\bigg)%\equiv V^{(0)}(\chi)+V^{(1)}(\chi)
\label{eq:curve40}\,,}
with effective masses
\ee{\mathcal{M}^2_\phi\equiv m^2+3\lambda\phi^2+\left(\xi- \f{1}{6}\right)R\,,\qquad \mathcal{M}^2_\psi\equiv g^2\phi^2+R/12\,.\label{eq:effm20}}
For this theory to one loop order the Callan-Symanzik equation (\ref{CZforV}) gives 
\ea{g^2(s)&=\f{g^2_0}{1-\f{5g^2_0}{8\pi^2}s} \\\f{\lambda(s)}{g^2(s)}&=\frac{-(10- 8N_{\rm f})\lambda_0-16N_{\rm f} g_0^2+ f(s)\left[8 N_{\rm f} g^2_0+\left(5-4N_{\rm f}+\sqrt{(1+4N_{\rm f})(25+4N_{\rm f})}\right)\lambda_0\right] }{(10- 8N_{\rm f})g_0^2-36\lambda_0+f(s)\left[18 \lambda_0+\left(-5+4N_{\rm f}+\sqrt{(1+4N_{\rm f})(25+4N_{\rm f})}\right)g^2_0\right]}  \\ 
%m^2(t)&=m^2_0\exp\bigg\{(16\pi^2)^{-1}\int_{0}^t\left(6\lambda(t) + 4g^2(t)\right) \dd t\bigg\}		\nonumber \\
\xi(s)&=\left(\xi_0-\f{1}{6}\right)\exp\bigg\{\f{1}{16\pi^2}\int_{0}^s\left[ 6\lambda(s') + 4N_{\rm f}g^2(s')\right] \dd  s'\bigg\}+\f{1}{6} \,,%\\
%m(t)&=0\,,
}
where in the above we use the notation
\ee{s\equiv \ln \left( \f{\mu}{\mu_0} \right)\,;\quad f(s)\equiv \bigg(\f{g_0}{g(s)}\bigg)^{\f{2}{5}\sqrt{(1+4N_{\rm f})(25+4N_{\rm f})}}+1\,,}
and where for clarity we have not written the analytic solution to the integral for $\xi$ as the result is very lengthy. For the purely gravitational operators one has
\ea{{V_\Lambda}&={\kappa}={\rm const.}\,\\
\alpha_1&=\f{1}{16\pi^2}\int_{0}^s\left\{ \f{1}{2}{\left[ \xi (s')-\f{1}{6}\right]^2-\f{N_{\rm f}}{72}}\right\} \dd  s'+\alpha_{1,0}\,\\
\alpha_2&=\f{4N_{\rm f}-1}{16\pi^2}\f{s}{180}+\alpha_{2,0}\,\\
\alpha_3&=\f{7N_{\rm f}+2}{16\pi^2}\f{s}{360}+\alpha_{3,0}\,,}
where the special case of de Sitter space comes via (\ref{eq:alp}).

\clearpage
\bibliographystyle{JHEP}
\bibliography{RGstoch_2}

\end{document}